\documentclass{article}

\usepackage{graphicx}
\usepackage{natbib}
\usepackage{multirow}
\usepackage{rotating}
\usepackage{bm}

\title{Adaptive hybrid optimization strategy for calibration and parameter estimation of physical models}
\author{Velimir V. Vesselinov (vvv@.lanl.gov) and Dylan R. Harp\\Computational Earth Sciences Group\\Earth and Environmental Science Division\\Los Alamos National Laboratory, Los Alamos, USA.}

\begin{document}
\maketitle

\begin{abstract}
A new adaptive hybrid optimization strategy, entitled \emph{squads}, is proposed for complex inverse analysis of computationally intensive physical models. Typically, models are calibrated and model parameters are estimated by minimization of the discrepancy between model simulations characterizing the system and existing observations requiring a substantial number of model evaluations. The new strategy is designed to be computationally efficient and robust in identification of the global optimum (e.g.\ maximum or minimum value of an objective function). It integrates a global Adaptive Particle Swarm Optimization (APSO) strategy with a local Levenberg-Marquardt (LM) optimization strategy using adaptive rules based on runtime performance. The global strategy optimizes the location of a set of solutions (particles) in the parameter space. The LM strategy is applied only to a subset of the particles at different stages of the optimization based on the adaptive rules. After the LM adjustment of the subset of particle positions, the updated particles are returned to the APSO strategy. Therefore, \emph{squads} is a global strategy that utilizes a local optimization speedup. The advantages of coupling APSO and LM in the manner implemented in \emph{squads} is demonstrated by comparisons of \emph{squads} performance against Levenberg-Marquardt (LM), Particle Swarm Optimization (PSO), Adaptive Particle Swarm Optimization (APSO; the TRIBES strategy), and an existing hybrid optimization strategy (hPSO). All the strategies are tested on 2D, 5D and 10D Rosenbrock and Griewank polynomial test functions and a synthetic hydrogeologic application to identify the source of a contaminant plume in an aquifer. Tests are performed using a series of runs with random initial guesses for the estimated (function/model) parameters. The performance of the strategies are compared based on their robustness, defined as the percentage of runs that identify the global optimum, and their efficiency, quantified by a statistical representation of the number of function evaluations performed prior to identification of the global optimum. \emph{Squads} is observed to have the best performance when both robustness and efficiency are taken into consideration than the other strategies for all test functions and the hydrogeologic application. 
\end{abstract}

\section{Introduction}
\label{Introduction}

Models are often used in the geosciences to indirectly estimate unknown (not observable) physical properties of a system based on observable quantities representing system behavior \citep{Carrera86,Dahlin01,Jessell01,Meek01,Poeter05}. In this process, the mathematical model is designed to simulate the system behavior $f(\bm{\theta})$ for a given set of model parameters $\bm{\theta}$ representing the actual physical properties of the system. The more accurately the model matches the observations, the more representative the model parameters are assumed to be. The process of making inferences about model parameters, commonly referred to as inverse modeling, regularly results in difficult optimization problems where a set of model parameters capable of acceptable representation of system behavior is sought. The optimization process is based on a metric representing the discrepancy between the model simulations $f(\bm{\theta})$ and the system observations. The discrepancy metric is also called the objective function (OF; $\Phi(\bm{\theta})$), and is a function of model parameters $\bm{\theta}$. In the parameter space, the metric is represented by a multi-dimensional response hyper-surface; a three-dimensional surface for the case of two model parameters. The response surface typically has a complex shape due to multiple minima (representing multiple plausible solutions) and flat regions (representing insensitivity of model parameters to the OF). An optimization process is based on a series of guided model evaluations for different model parameter sets. The challenges in the optimization process come from complications in identifying the global minima and from requirements to execute a substantial number of model evaluations. Frequently, the number of model evaluations needed for optimization can vary from about 100 to more than $10^{6}$ depending on the complexity of the inverse model. As a result, the optimization process can be especially difficult in real-world applications using physical models where a single forward model simulation is performed from several minutes to more than an hour. In these situations, even efficient parallel techniques (e.g.\ \cite{Vesselinov01}) can cause substantial computational burden. Therefore it is important to develop computationally efficient and robust strategies that can identify the global minimum with a relatively small number of model evaluations.

Optimization strategies can be classified as global and local strategies \citep{Nocedal99}. Global strategies excel at robust exploration of the response surface, identifying multiple areas of attraction; however, global strategies are inefficient at locating the parameter set producing an optimal solution within an area of attraction. As a result, in the case of real world model inversions, the application of global strategies may be infeasible when the model evaluations take a substantial amount of computational time \citep{Keating10}. Local strategies excel at efficiently identifying the optimal model parameters within an area of attraction; however, local strategies are not designed for robust exploration of a response surface outside of an area of attraction. The local strategies are efficient within an area of attraction because they utilize local information about the gradient and curvature of the response surface. This requires estimation of the first and second order derivatives of the discrepancy metric in the parameter space. As many real world model problems have response surfaces with multiple minima, the use of local strategies alone is not always robust. One of the most commonly used local strategies is the Levenberg-Marquardt (LM) strategy which has been applied in many frequently used inverse analysis and parameter estimation codes in the geosciences such as UCODE \citep{Poeter99} and PEST \citep{Pest05}.

Global and local strategies are complimentary; where one excels, the other struggles, and vice versa. The benefits of hybrid global/local strategies have been demonstrated previously \citep{Noel04,hPSO,Zhang07,Ghaffari-Miab07}. We introduce a new development in hybrid optimization, coupling recent developments in Adaptive Particle Swarm Optimization (APSO) and a Levenberg-Marquardt (LM) strategy producing a novel adaptive hybrid strategy entitled \emph{squads}. The strategy applies an LM strategy to a subset of particles at different stages of an APSO strategy based on adaptive rules. After the LM update of the particle position, the particle is passed back to the APSO strategy and continues to evolve based on APSO rules. In essence, \emph{squads} is a global strategy utilizing local optimization speedup. \emph{Squads} is specifically designed to be a robust and computationally efficient strategy capable of identifying the global minimum with a relatively small number of model evaluations in complex inverse problems. The name \emph{squads} refers to the hierarchical structure of the population of solutions in the algorithm, similar to the APSO algorithm TRIBES \citep{tribes}.

The \emph{squads} strategy is tested using the Rosenbrock \citep{Rosenbrock60} and Griewank \citep{Griewank81} polynomial test functions. \emph{Squads} is also applied to solve a hydrogeological problem related to identification of the source of a contaminant plume in an aquifer; this problem is frequently solved in applications related to protection and remediation of groundwater resources \citep{Bagtzoglou91,Snodgrass97,Atmadja01,Sun06,Dokou09}. In order to demonstrate the relative benefits of the hybrid strategy of \emph{squads}, its performance is compared to open-source distributions of LM \citep{lourakis04LM}, PSO \citep{pso06}, and APSO \citep{tribes} strategies. Additionally, \emph{squads} is compared to hPSO \citep{hPSO}, an open-source hybrid strategy that combines PSO and the Nelder-Mead downhill simplex strategy \citep{Nelder65} implemented in the MATLAB$^{\mbox{\textregistered}}$ \citep{matlab} computing environment.

\section{Particle Swarm Optimization}

Sociobiologists have theorized that individuals within a population can benefit from the previous knowledge and experience of other members of the population while searching for sporadically distributed food sources \citep{Wilson75}. The ubiquity of schooling and flocking tendencies common among many species suggests that this is an efficient, cost-effective strategy for the survival of individuals. It is easy to recognize the analogy of organisms searching for food sources and mathematical strategies searching for optimal solutions. This recognition led to the development of PSO by \cite{Kennedy95}, building on previous research intended to graphically simulate the flocking behavior of birds. Certain aspects of the flocking behavior of this early research has been eliminated in order to improve the strategy's performance in global optimization, leading to the use of the term ``swarm'' to describe the graphical behavior of PSO.

The development of PSO has produced a parsimonious optimization strategy modeling a population of randomly selected initial solutions (particles) by their position and velocity \citep{Clerc06}. In a $D$-dimensional parameter space, the position and velocity of the $i$th particle can be represented as $\mathbf{p}_i = [p_{i,1},p_{i,2},\dots,p_{i,D}]$ and $\mathbf{v}_i = [v_{i,1}, v_{i,2},\ldots,v_{i,D}]$, respectively. An empirical formula for determining the swarm size $S$ has been suggested as $S=10+2\sqrt{D}$ \citep{pso06}. Particles retain a record of the best location they have visited so far denoted as $\mathbf{b}_i = [b_{i,1}, b_{i,2},\ldots,b_{i,D}]$. Particles are also informed of the best location that $K$ other randomly chosen particles have visited, denoted as $\mathbf{g}_i = [g_{i,1}, g_{i,2},\ldots,g_{i,D}]$. A standard value for $K$ is 3 \citep{pso06}. These networks of informers are reinitialized after iterations with no improvement to the global best particles of the swarm. The velocity of the $i$th particle in the $j$th dimension is updated from strategy iteration $k$ to $k+1$ as

\begin{equation} 
	v_{i,j}(k+1) = w v_{i,j}(k) + c_1 r_1 (b_{i,j} - p_{i,j}(k)) + c_2 r_2 (g_{i,j} - p_{i,j}(k)), \quad k = \{1,\dots,D\},
	\label{eq:vel}
\end{equation}

\noindent where $w$ is a constant referred to as the inertia weight, $c_1$ and $c_2$ are constants referred to as acceleration coefficients, $r_1$ and $r_2$ are independent uniform random numbers in $[0,1]$. The parameter $w$ controls the level of influence between its previous and current particle displacement, $c_1$ and $c_2$ scale the random influence of (1) the particle memory (past particle locations in the parameter space), and (2) the current network of particle informers (current informer locations in the parameter space), respectively. A limitation on the magnitude of the velocity $V_{max}$ is commonly employed. The particle position is updated during each strategy iteration as

\begin{equation}
	p_{i,j}(k+1) = p_{i,j}(k) + v_{i,j}(k+1),\quad k = \{1,\dots,D\}.
\end{equation}

It has been recognized that the selection of $w$, $c_1$, $c_2$, and $V_{max}$ tune the performance of PSO, modifying the balance between exploration (spreading the particles throughout the parameter space) and intensification (focusing the particles within an area of attraction). Manual tuning of PSO's parameters can be a delicate task. Adaptive PSO (APSO) strategies have emerged in order to reduce or eliminate the often difficult and time-consuming process of parameter tuning of PSO \citep{Cooren09}. 

One of the algorithmic variants of APSO is TRIBES \citep{Clerc06} (TRIBES is not an acronym, but we follow the convention of all capital letters as proposed by its designer), which eliminates the tuning of the PSO strategy parameters. The strategy has been proven competitive with well-known strategies on a suite of test problems \citep{Cooren09}. As the name suggests, TRIBES partitions the particles into groups, referred to as \emph{tribes}, intended to facilitate the exploration of multiple areas of attraction. In this way, a hierarchical structure is established where the swarm is composed of a network of tribes, and each tribe is a network of particles. The intent is to eliminate parameter tuning as the swarm evolves from an initial set of tribes, and the tribes evolve from single particles based on rules governing the evolution of the swarm topology and rules for generation and elimination of entire tribes and individual particles within the tribes. The particle within a tribe with the lowest/highest OF for minimization/maximization is considered the shaman of the tribe. Information is shared only between the particles within a given tribe. Information between the tribes is shared only through the shamans. In this way, the displacements of non-shaman particles are influenced by the shaman of their tribe, while the displacements of the shamans are influenced by the \emph{best} shaman in the swarm. The source code for TRIBES is available from \cite{tribes}.  

\section{\emph{Squads} adaptive hybrid optimization}
\label{sect:squads}

Various approaches have been introduced to couple the global search capabilities of PSO with the efficiency of first and second-order local strategies. \cite{Clerc99} introduced a PSO strategy that adjusts particle locations based on approximations of the gradient of the OF utilizing the OF values of the current particle locations. \cite{Noel04} developed a hybrid PSO strategy incorporating gradient information directly in the calculation of particle velocity. \cite{hPSO} coupled a PSO strategy with the Nelder-Mead simplex strategy (hPSO, \cite{Lagarias98}), \cite{Zhang07} coupled PSO and back-propagation to train neural networks. \cite{Ghaffari-Miab07} developed a hybrid strategy, iterating between PSO and BFGS quasi-Newton optimization. We present a hybrid strategy called \emph{squads} that couples an APSO strategy with a Levenberg-Marquardt (LM) strategy. The following provides a detailed description of a fine-tuned coupling of APSO and LM based on adaptive rules, where the LM optimization is applied to improve the locations of a subset of selected particles (the shamans) in the course of the optimization process. The current APSO strategy implemented in \emph{squads} is TRIBES \citep{Clerc06}, and the LM optimization is performed using the LevMar library \citep{lourakis04LM}.

Much of the time-consuming and difficult tuning required of many optimization strategies is reduced in \emph{squads} utilizing adaptive rules. The APSO strategy does not require the specification of optimization parameters \citep{Clerc06}, and the applied LM strategy is optimized to work well on many problems using default and internally estimated optimization parameters \citep{lourakis04LM}. The adaptive rules implemented in \emph{squads} to control the performance of LM speedups during the APSO optimization are also designed to be general and capable to tackle problems with different complexity.

\begin{figure} \begin{center}
		\includegraphics[width=6cm]{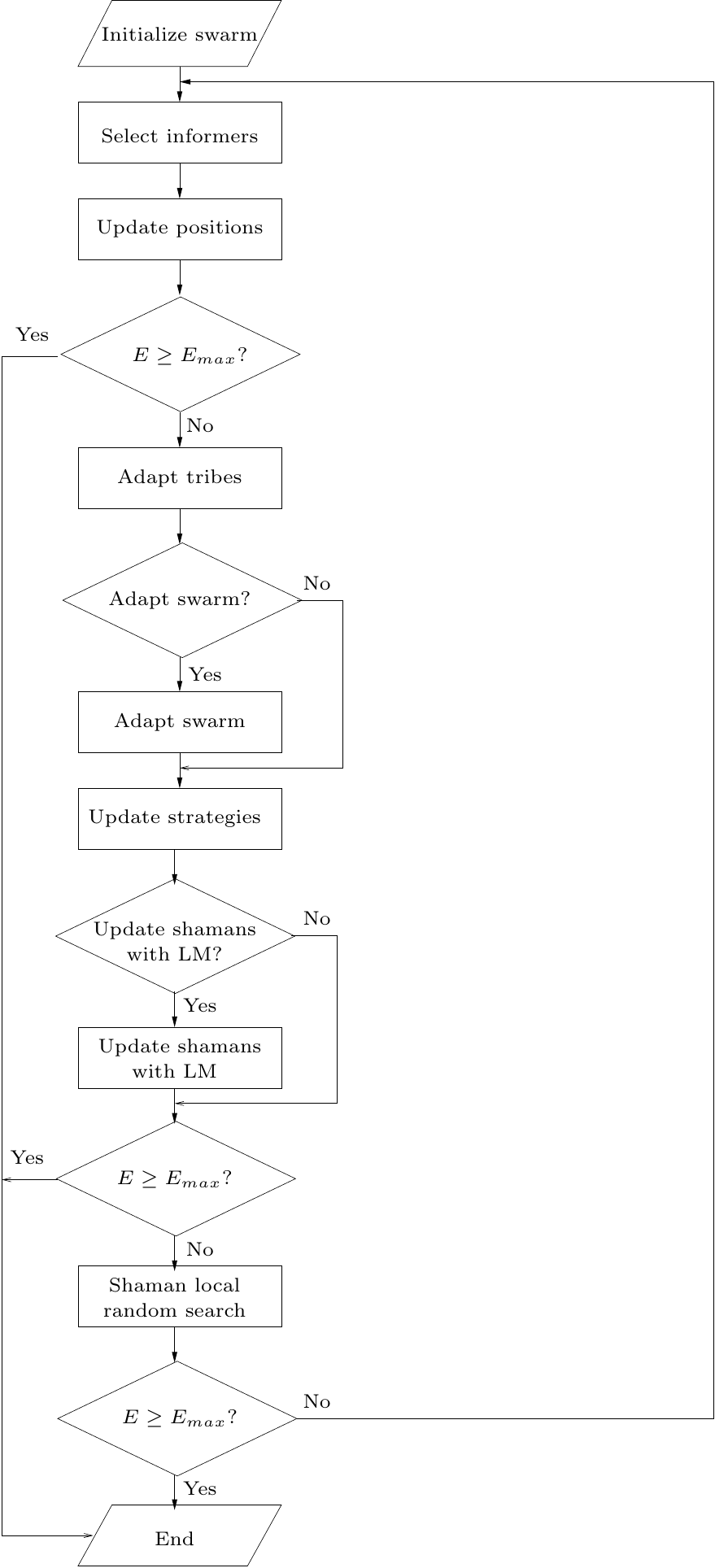}
		\label{fig:flow}
		\caption{Flow diagram of \emph{squads} strategy. $E$ is the current number of model evaluations and $E_{max}$ is the allowable number of model evaluations. Decisions to ``adapt swarm'' or ``update shamans with LM'' are determined by adaptive rules.}
	\end{center}
\end{figure}

A flow diagram of the \emph{squads} strategy is presented in Figure~\ref{fig:flow}. Tables~\ref{tab:init} and \ref{tab:disp} describe the particle initialization and displacement rules and their selection within \emph{squads}. For consistency with other global strategies discussed here, \emph{squads} is initialized with $N_t=S=10+2\sqrt{D}$ mono-particle tribes, where $N_t$ is the number of tribes in the swarm and $S$ is the number of particles. However, \emph{squads} can also be initiated with a single mono-particle tribe and allow the swarm to develop based on the built-in adaptive rules. If provided, one of the initial particles is set to predefined values (rule 1\ in Table~\ref{tab:init}), while the remaining positions of the  initial particles are determined according to rule 5\ in Table~\ref{tab:init}. 

\begin{table}
	\caption{Particle initialization rules and their selection criteria.}
\ \\Particle initialization rules:
	\begin{enumerate}
		\item User specified
		\item Randomly chosen position within parameter space:\\
			$p_{new_j} = $U$(p_{min_j}, p_{max_j}), \quad j = {1,\ldots,D}$
		\item Randomly chosen within hyperparallelepid surrounding the best position of the swarm with dimensions ($2\cdot r_j$) determined by Euclidean distance between the swarm's and tribe's best position:\\
			$r_j = |p_{best_j} - p_{tribe\ best_j}| \quad j = {1,\ldots,D}$ \\
			$p_{new_j} = $U$( p_{best_j} - r_j, p_{best_j} + r_j ) j = {1,\ldots,D}$
		\item On one of the vertices of the parameter space with equal probability of being the max or min of each dimension:\\
			$\mbox{if }($U$(0,1) < 0.5) \mbox{ then } p_{new_j} = p_{min_j} \mbox{, else } p_{new_j} = p_{max_j}\quad j = {1,\ldots,D}$
		\item Randomly chosen within the largest empty hyperparallelepid of the parameter space
	\end{enumerate}
	\begin{center}
	\begin{tabular}{l|l}
	Criteria&Initialization rule selection\\
	\hline
	First particle of the strategy&1\\
	\hline
	If initial population is greater&\multirow{2}{*}{5}\\
	than 1, other initial particles&\\
	\hline
	Particle added to ``bad'' tribe (tribe adaptation)&randomly chosen between 2 and 5\\
	\hline
	Mono-particle tribe added (swarm adaptation)&5\\
	\hline
	LM unable to reduce OF of shaman by 2/3&5\\
	\end{tabular}	
	\label{tab:init}
	\end{center}
\end{table}

\begin{table}
	\caption{Particle displacement rules and their selection criteria based on the status of the particle. N$(\mu,\sigma)$ is a normal distribution with a mean $\mu$ and standard deviation $\sigma$, U$(a,b)$ is a uniform distribution with minimum $a$ and maximum $b$, $f(-)$ is the value of the objective function, $\mathbf{g}=[g_1,g_2,\dots,g_D]$ is the location of the particles designated informer, $\mathbf{b}=[b_1,b_2,\dots,b_D]$ the particle's current best location, and $min_j$ and $max_j$ are the minimum and maximum values for the $j$th dimension, respectively.}
\ \\Particle displacement rules:
	\begin{enumerate}
		\item $p_{j} = $U$(min_j,max_j)\quad j = {1,\ldots,D}$, change displacement rule to 2 for next time
		\item $p_{j} = $N$(g_j, 0.74\cdot| b_j - g_j |)$\\
					or, if no informer\\	
					$p_{j} = $N$(b_j, \max(b_j-min_j,max_j-b_j))$
		\item $p_{j} = \frac{f(\mathbf{g})}{f(\mathbf{g})+f(\mathbf{b})}\cdot\mbox{U}(b_j - | b_j - g_j |, b_j + | b_j - g_j |) + \frac{f(\mathbf{b})}{f(\mathbf{g})+f(\mathbf{b})}\cdot\mbox{U}(g_j - | b_j - g_j |, g_j + | b_j - g_j |)$\\
					or, if no informer\\
					$p_{j} = $N$(b_j, 3\cdot\max(b_j-min_j,max_j-b_j))$
	\end{enumerate}
	\begin{center}
	\begin{tabular}{c|l}
	Particle status&Displacement rule selection\\
	\hline
	(--=)&randomly choose any rule other than current one\\
	(==)&randomly choose between rule 2 and 3\\
	(+=) or (++)&change to rule 1 with 50\% probability\\
	\end{tabular}	
	\label{tab:disp}
	\end{center}
\end{table}

Each iteration of the strategy is initiated by determining the informers for all the particles. For non-shaman particles, this will be the shaman of their tribe. A shaman is the particle with the best (e.g.\ lowest for minimization) OF value within the tribe. For shaman's, this will be the shaman with the best OF value within the swarm, referred to as the \emph{best shaman}. Particle positions are then updated according to the rules described in Table~\ref{tab:disp}. Particles are initialized to use displacement rule 1. After informers are determined, particle positions are updated based on their currently selected displacement rule.

The decision to adapt a tribe is based on whether the tribe has demonstrated sufficient improvement during the previous strategy iteration. This is performed stochastically, by comparing the fraction of particles in the tribe that improved their location in the last move with a random number between 0 and 1. If the fraction is greater than the random number, the tribe is considered a \emph{good tribe}, and the worst particle is removed from the tribe. This eliminates unnecessary model evaluations, focusing the attention of the tribe on the \emph{good particles}. Otherwise, the tribe is considered a \emph{bad tribe}, and a particle is added to the tribe (refer to Table~\ref{tab:init} for details on particle initialization rule selection) and a randomly selected dimension of a randomly selected particle in the tribe (other than the shaman) is reinitialized randomly. Adding a particle to a \emph{bad tribe} is intended to increase the exploration of the parameter space by the tribe. 

The swarm adaptation occurs either every $N_t*(N_t-1)/4$ strategy iterations or if the swarm is labeled by LM as a \emph{bad swarm}. A swarm is considered a \emph{bad swarm} if LM speedup was performed in the previous iteration, and the OF was not reduced by at least 2/3 for all the LM updated shamans. A mono-particle tribe is added to the swarm if it is considered \emph{bad} according to rule 5\ in Table~\ref{tab:init}. The tribe led by a shaman with the worst OF in the swarm is removed if the swarm is considered \emph{good}.

Next, particle displacement rule selections are updated. Particle displacement rule selections are modified based on whether or not (1) their position has improved in the last move and (2) their best overall position has improved in the last move. Following the convention of \cite{Clerc06}, we use a (+) to indicate improvement, (=) the same OF value, and (--) a worse position. The particles performance can then be denoted as one of the following: (--=), (==), (+=), and (++), where the first symbol indicates if the particle improved its position in the last move, and second symbol indicates if the overall best position of the particle improved in the last move. Note that the best overall performance can only stay the same or improve, and an improvement in the overall performance indicates an improvement over the last position. Table~\ref{tab:disp} lists the displacement rule selection based on particle performance.

After the swarm adaptation, \emph{squads} checks whether or not to update the shamans using LM. LM updating is turned off in \emph{squads} if none of the shamans reduces the OF of the previous shamans by more than 2/3 during the last LM updating. LM updating will be restarted when the best OF of the previously obtained OF during LM has been reduced by an order of magnitude by the APSO strategy. This postpones LM until the global APSO strategy has identified a position with a significant improvement, which will perhaps be a previously unidentified area of attraction. After the LM optimization, the new shaman location is used in the APSO strategy. 

In contrast with the APSO strategy, the LM strategy requires that the OF be represented as a summation of components at least equal to the number of parameters as 

\begin{equation}
	\Phi(\bm{\theta}) = \sum_{i=1}^{N}\Phi_i(\bm{\theta}),
\label{eq:Phi}
\end{equation}

\noindent where $\bm{\theta}$ is a vector of model parameters and $N$ is equal or larger than the number of model parameters. This allows the LM strategy to estimate the local gradient and curvature of the response surface in the parameter space. These calculations utilize numerical derivatives of the OF components in equation~\ref{eq:Phi} with respect to the model parameters (also called the Jacobian matrix). Based on the Jacobian matrix, the LM strategy also estimates the second-order derivatives of the OF components with respect to model parameters (also called a Hessian matrix). The second-order derivatives approximate the local curvature of the response surface. The LM strategy searches for the local optimum by adaptive adjustment between first and second-order optimization techniques \citep{Levenberg44,Marquardt63}. Frequently in the case of model inversion problems, the OF in equation~\ref{eq:Phi} is represented by the discrepancy between model simulated values $f_i(\bm{\theta})$ and corresponding observations $o_i$, where $i = 1, ..., N$, and $N$ is now the number of observations. For example, frequently $\Phi(\bm{\theta})$ is computed as 

\begin{equation}
	\Phi(\bm{\theta}) = \sum_{i=1}^{N}\Phi_i(\bm{\theta}) = \sum_{i=1}^{N} (f_i(\bm{\theta}) - o_i)^2.
\end{equation}

\noindent \emph{Squads} estimates the first-order derivatives using a finite difference approach applied in the LevMar library \citep{lourakis04LM}. 

The following criteria are defined by default in LevMar to terminate the LM optimization \citep{lourakis04LM}, and applied in the LM updating of \emph{squads} as well: (1) the maximum change in any parameter is less than $10^{-5}$; (2) the relative change in the L2 norm of the change in the parameter values is less than $10^{-5}$ of the L2 norm of the parameter values; (3) the OF reaches a value of zero; (4) the Jacobian matrix is close to singular, and (5) the maximum number of LM iterations (i.e.\ derivative approximation and Marquardt parameter value exploration) is achieved (50 when standalone LM is performed; 8\ in \emph{squads}). The criteria are designed to terminate LM once it successfully identifies a local optimum. Typically, criteria 1, 2, and 5 terminate the LM updating in \emph{squads} (the termination criteria of the LM updating within \emph{squads} do not terminate the \emph{squads} run). \emph{Squads} is terminated when either one of the following conditions are met: (1) $E_{max}$, the number of allowable model evaluations, is exceeded or (2) the OF reaches below a predefined cutoff value. 

The final step of each iteration is to perform a random local search in the empty space around each shaman \citep{tribes}. In this step, a random position within the largest hyperparallelepid centered on the tribe's shaman, void of other particles, is evaluated. If the position is an improvement over the current shaman position, the shaman is moved to this location. Otherwise, the position is forgotten.

Global strategies in general, including APSO, are designed to operate on a bounded parameter space. The parameter ranges are typically predefined depending on the physical constraints or prior knowledge about the parameter distributions. However, the LM optimization by default works in an unbounded parameter space. There are various techniques to constrain an LM strategy within a parameter space, but these techniques typically have a negative impact on LM performance. To avoid this, \emph{squads} operates in a transformed parameter space. For example, an element of the parameter vector $\bm{\theta}$ is transformed as

\begin{equation}
	\hat{\theta} = \arcsin\left ( \frac{\theta - \theta_{min}}{\theta_{max} - \theta_{min}} \cdot 2 - 1 \right ),%\quad p_t \in left [-\frac{\pi}{2},\frac{\pi}{2}\right ],
\end{equation}

\noindent where $\hat{\theta}$ is the transformed parameter, and $\theta_{max}$ and $\theta_{min}$ are the upper and lower bounds for parameter $\theta$, respectively. The APSO strategy is performed in the transformed parameter space bounded within $[-\pi/2; \pi/2]$ in all dimensions, while the LM updating is performed unconstrained in the transformed parameter space. Model (function) evaluations are performed on de-transformed parameters by

\begin{equation}
	\theta = \theta_{min} + \left ( \frac{\sin(\hat{\theta}) + 1}{2} \right ) (\theta_{max} - \theta_{min}).
\end{equation}

\noindent In this way, the LM updating is unaware of parameter boundaries and is unaffected by performance issues associated with calculating numerical derivatives near boundaries. It should be noted that in the process of the LM updating, the transformed parameters can be moved outside of the $[-\pi/2; \pi/2]$ range; however, the transformed parameters are returned to equivalent values within $[-\pi/2; \pi/2]$ before being passed back to the APSO strategy by

\begin{equation}
	\hat{\theta}_{APSO} = \arcsin( \sin(\hat{\theta}_{LM}) ).
\end{equation}

\noindent where $\hat{\theta}_{LM}$ represents the unconstrained transformed parameters resulting from LM updating and $\hat{\theta}_{APSO}$ represents the constrained transformed parameters passed back to the APSO strategy, thereby ensuring that APSO receives parameters within its explicitly defined, bounded parameter space. It is important to note that $\hat{\theta}_{APSO}$ and $\hat{\theta}_{LM}$ are equivalent in the non-transformed parameter space.

\section{Test functions}

The \emph{squads} strategy is tested by optimizing the Rosenbrock and Griewank test functions. The Rosenbrock and Griewank functions present difficult optimization problems exhibiting frequently observed complexities in response surface topology in real world problems (e.g.\ \cite{Rosenbrock60,Griewank81,Clerc06,Cooren09}).

\begin{figure}
	\begin{center}
		\includegraphics[width=12cm]{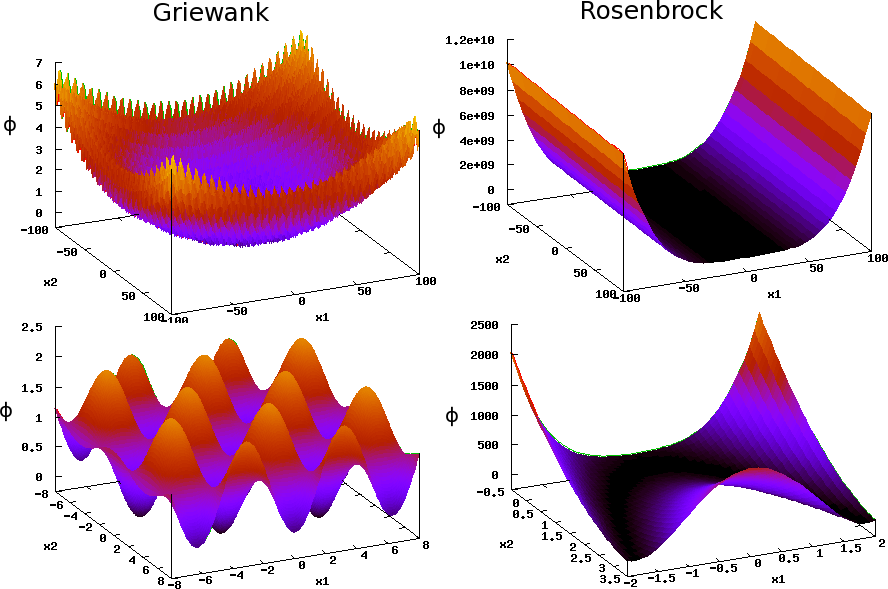}
		\caption{Rosenbrock and Griewank polynomial test functions with global minima at (1,1) and (0,0), respectively. Note the different parameter ranges on the top and bottom rows. The top row shows the parameter space explored by the optimization strategies. The bottom row focuses on the parameter space near the respective global minima.}
		\label{fig:tests_3D}
	\end{center}
\end{figure}

The response function defined by the Rosenbrock function is comprised of a large valley with an ill-defined, shallow global minimum. For $D\le3$, the function is unimodal with a global minimum at $\mathbf{x}=\mathbf{1}$ (where $\mathbf{1}=[1,\dots,1]$). For $4\le D \le 7$, a local minimum exists at $(x_1,x_2,\dots,x_D) = (-1,1,\dots,1)$ in addition to the global minimum, while for $D>7$, multiple suboptimal local minima exist \citep{Shang06}. In the case of two model parameters, the shape of the Rosenbrock function is presented in Figure~\ref{fig:tests_3D}. The Rosenbrock function generalized to any number of dimensions greater than or equal to two can be expressed as

\begin{equation}
	\Phi_r(x_1,\dots,x_D) = \sum_{i=1}^{D-1} (1-x_i)^2 + 100(x_{i+1}-x_i^2)^2.
\end{equation}

\noindent The estimation of the local gradient and curvature of the response surface by LM requires the test function to be represented as a summation of parts as in equation~\ref{eq:Phi}. The summation components of $\Phi_r(x_1,\dots,x_D)$ can be expressed as

\begin{equation}
	\Phi_{r,2i-1}(x_i) = (1-x_i)^2\ i < D
	\label{eq:phir1}
\end{equation}

\noindent and

\begin{equation}
	\Phi_{r,2i}(x_i,x_{i+1}) = 100(x_{i+1}-x_i^2)^2\ i < D
	\label{eq:phir2}
\end{equation}

\noindent producing $2(D-1)$ OF components where equation~\ref{eq:phir1} and \ref{eq:phir2} define the odd and even numbered components, respectively; therefore, the number of components (called also ``observations'' in the case inverse problems) in the 2D, 5D, and 10D cases are 2, 8, and 18, respectively. The LM strategy uses the derivatives of $\Phi_{r,i}(x_1,\dots,x_D)$ with respect to model parameters to evaluate the local gradient and curvature of the response surface. In most real world problems, the analytical computation of derivatives is not feasible. Therefore, in all the examples presented below, the derivatives are computed numerically using a finite difference approach, even though the analytical derivation in this case is trivial. Other alternative representations of $\Phi_r$ as a sum of components are also possible.

The $D$-dimensional Griewank function is defined as

\begin{equation}
	\Phi_g(x_1,\dots,x_D) = 1 + \frac{1}{4000} \sum_{i=1}^D x_i^2 - \prod_{i=1}^D \cos\left (\frac{x_i}{\sqrt{i}}\right ).
\end{equation}

\noindent The Griewank function has numerous local areas of attraction, but a single global minimum of zero at $\mathbf{x} = \mathbf{0}$. In the two-dimensional case, the function has the shape of an ``egg carton'' that is depressed in the center, as depicted in Figure~\ref{fig:tests_3D}).

The summation components can be defined as

\begin{equation}
	\Phi_{g,i}(x_1,\dots,x_D) =  \frac{1}{D} + \frac{x_i^2}{4000} - \frac{1}{D} \prod_{i=1}^D \cos \left(\frac{x_i}{\sqrt{i}}\right)
\end{equation}

Therefore, the number of components (``observations'') equals the number of model parameters.

The multidimensional Griewank function is important for testing of hybrid optimization strategies because it becomes more difficult to minimize for global strategies as its dimensionality increases \citep{Locatelli03}. However, although counterintuitive, the Griewank function becomes easier to minimize for local strategies as the dimensionality increases. Therefore, with the increase in dimensionality, it is expected that LM performance will improve while the PSO, TRIBES and hPSO performance will decrease. For different parameter-space dimensionality, the performance of hybrid strategies will depend on how efficiently they adaptively balance between the local and global strategies. At low dimensionality ($D=2$), the hybrid strategies should benefit from the global strategy; at high dimensionality, the hybrid strategies should benefit from the local strategy.

\section{Contaminant source identification test case}
\label{sect:cont}

\begin{figure}
	\begin{center}
		\includegraphics[width=8cm]{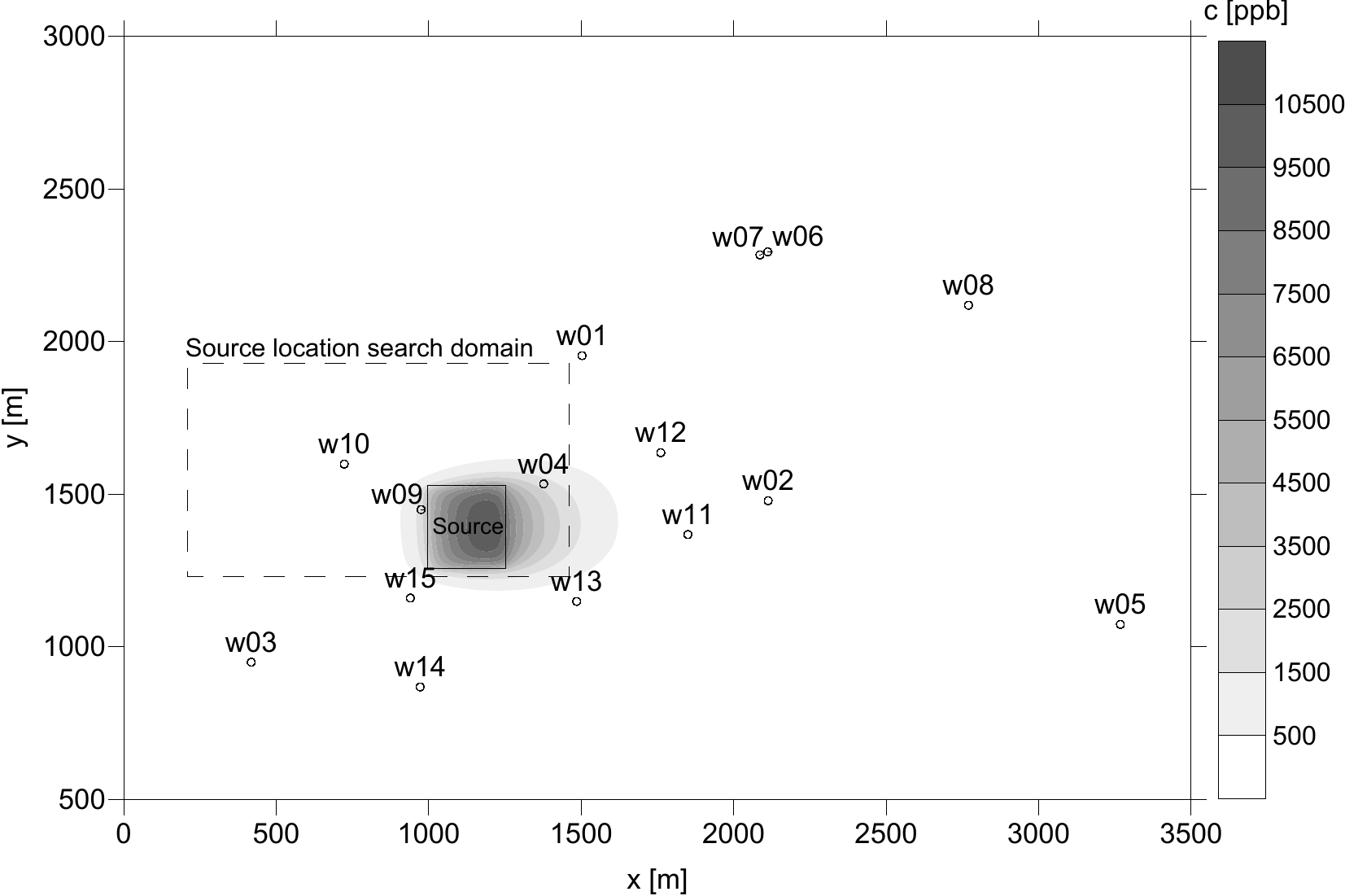}
		\caption{Map of monitoring well locations. The ``true'' source is shown as a solid rectangle. The search domain for $x_s$ and $y_s$ is shown as a dotted rectangle. The contaminant concentration plume at $t=49$ years is represented by the color map.}
		\label{fig:map_sid}
	\end{center}
\end{figure}

Optimization strategies are commonly employed to calibrate physics-based models to available observations. We demonstrate the optimization strategies on a hydrogeologic application to identify the center ($x_s, y_s$) and dimensions ($x_d, y_d$) of a parallelepiped contaminant source in an aquifer using observations of contaminant concentrations from monitoring wells near the expected source location. The synthetic groundwater flow and transport problem is three-dimensional and semi-infinite, the top model boundary aligns with the top of the aquifer, and the model extends to infinity laterally and with depth. The locations of the monitoring wells, depths below the aquifer top boundary of the top and bottom of the screens, times of observation after the contaminant release, and observed contaminant concentrations are presented in Table~\ref{tab:targets}. The parameter values used to generate the ``true'' concentrations at the monitoring wells and the minimum and maximum parameter values allowed in optimization runs are presented in Table~\ref{tab:pars}. The ``true'' location of the source, the location of monitoring wells, and contaminant concentrations at $t=49$~years since the contaminant was released are presented in Figure~\ref{fig:map_sid}. A similar model is presented in \cite{Harp11} with additional details.

\begin{table}
        \begin{center}
                \begin{tabular}{c|c|c|c|c|c|c}
                Well&$x$ [m]&$y$ [m]&$z_{top}$ [m]&$z_{bot}$ [m]&$t$ [a]&$c$ [ppb]\\
                \hline
		w01 &1503 &1954 &5.57 &12.55 &49 &0\\
		w02 &2113 &1479 &36.73 &55.14 &49 &0\\
		w03 &418 &950 &0 &15.04 &49 &0\\
		\hline
		\multirow{2}{*}{w04} &\multirow{2}{*}{1377} &\multirow{2}{*}{1534} &\multirow{2}{*}{13.15} &\multirow{2}{*}{20.41} &44 &350\\
		&&&&&49 &432\\
		\hline
		w05 &3268 &1074 &26.73 &33.71 &49 &0\\
		w06 &2112 &2294 &69.01 &83.98 &49 &0\\
		w07 &2086 &2284 &11.15 &18.19 &49 &0\\
		w08 &2770 &2119 &4.86 &11.87 &49 &0\\
		w09 &975 &1450 &3.66 &10.09 &49 &981\\
		\hline
		\multirow{2}{*}{w10} &\multirow{2}{*}{723} &\multirow{2}{*}{1599} &3.32 &9.63 &49 &1.1\\
		&&&23.2 &26.24 &49 &0.1\\
		\hline
		\multirow{2}{*}{w11} &\multirow{2}{*}{1850} &\multirow{2}{*}{1368} &4.94 &7.99 &49 &22\\
		&&&32.46 &35.48 &49 & 0.3\\
		\hline
		\multirow{2}{*}{w12} &\multirow{2}{*}{1761} &\multirow{2}{*}{1636} &3.59 &6.64 &49 &15\\
		&&&32.51 &38.61 &49& 0.17\\
		\hline
		\multirow{2}{*}{w13} &\multirow{2}{*}{1485} &\multirow{2}{*}{1149} &3 &6 &50 &72\\
		&&&36 &42 &50& 0.26\\
		\hline
		w14 &972 &869 &3 &6 &50 &0\\
		w15 &940 &1160 &3 &6 &50 &38\\
                \end{tabular}
                \label{tab:targets}
        \caption{Well coordinates, screen top ($z_{top}$) and bottom ($z_{bot}$) depths below the water table, and year and  value of observed contaminant concentrations.}
        \end{center}
\end{table}

\begin{table}
        \begin{center}
                \begin{tabular}{l|c|c|c|c}
                &$x_s [m]$&$y_s [m]$&$x_d [m]$&$y_d [m]$\\
                \hline
								`true'&1124&1393&258&273\\
								min&210&1230&1&1\\
								max&1460&1930&500&500\\
                \end{tabular}
                \label{tab:pars}
        \caption{`True', minimum, and maximum parameter values for the contaminant transport test case.}
        \end{center}
\end{table}

The objective function for the contaminant transport test case is a sum-of-the-squared-residuals (SSR) expressed as

\begin{equation}
  \Phi(\bm{\theta}) = \displaystyle \sum_{i=1}^N (\hat{c}_i(\bm{\theta}) - c_i)^2,
	\label{eq:phi-cont}
\end{equation}

\noindent where $\hat{c}_i(\theta)$ is the $i^{\mbox{th}}$ simulated concentration resulting from $\bm{\theta}$, $c_i$ is the $i^{\mbox{th}}$ observed concentration, and $N$ is the number of observations. In summary, there are 4 unknown model parameters constrained by 20 observations.

The simulated contaminant concentrations ($\hat{c}$) are produced from an analytical contaminant transport model encoded in MADS \citep{Wexler92,Wang09,MADS10} (refer to \cite{Harp11} for additional simulation details). Due to the rounding of the observed concentrations, a value of $\Phi$=0.55 is obtained from the true parameter values.

\begin{figure}
	\begin{center}
		\includegraphics[width=8cm]{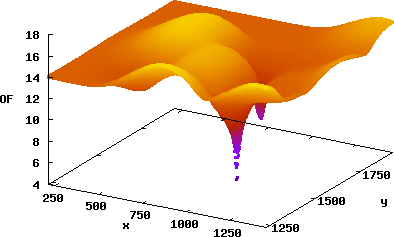}
		\caption{Contaminant source identification response surface for contaminant source locations defined by $x_s$ and $y_s$. The minimum OF value for each combination of $x_s$ and $y_s$ are plotted considering allowable ranges for source lateral dimensions, $x_d$ and $y_d$.}
		\label{fig:cont_surf}
	\end{center}
\end{figure}

The response surface resulting from equation~\ref{eq:phi-cont} plotted as a function of $x_s$ and $y_s$ is presented in Figure~\ref{fig:cont_surf}. The values plotted in Figure~\ref{fig:cont_surf} are the lowest values for $\Phi$ at each combination of $x_s$ and $y_s$ considering allowable ranges for $x_d$ and $y_d$. In other words, this is the response surface that the strategies would traverse if they knew the optimal values for $x_d$ and $y_d$ for each combination of $x_s$ and $y_s$. Note that the actual 4D response surface is more complicated than this 2D representation. Features from both the Griewank and Rosenbrock test functions can be seen in this representation of the response surface with multiple areas of attraction (three suboptimal minima and one global minimum), regions of parameter insensitivity (flat regions), and narrow, curved (banana-shaped) valleys.

Even though the contaminant transport model is analytical, the computational time is substantially higher than that for the test functions. Within MADS, the number of function evaluation per second is \~40,000 for the test functions compared to \~400 for the contaminant transport model. For hPSO, the number of function evaluations per second is \~11,000 for the test functions compared to \~2 for the contaminant transport model; the substantial increase in hPSO computation time between the test functions and the hydrogeologic application is due to external coupling between the Matlab computing environment (applied to execute hPSO) and external C based transport simulator.

\section{Results and discussion}

The performance of \emph{squads} on the Rosenbrock and Griewank functions is compared with (1) LM, (2) PSO, (3) TRIBES, and (4) hPSO. The LM strategy is an implementation of LevMar \citep{lourakis04LM} (the same strategy as applied in \emph{squads}), PSO is an implementation of Standard PSO 2006 \citep{pso06}, TRIBES is an APSO strategy implemented in the code described in \cite{Clerc06}, hPSO is freely available hybrid optimization code from \cite{hPSO}. LM, PSO, TRIBES and \emph{squads} are built into the code MADS \citep{MADS10}, which is utilized for all analyses except hPSO. The hPSO analysis is performed using MATLAB version 7.8.0.347 (R2009a) \citep{matlab}. The strategy parameters (i.e.\ optimization parameters) for PSO and hPSO are set to values that have been demonstrated to perform well in many test cases \citep{pso06} as $w=0.72$, $c_1=1.2$ and $c_2=1.2$ (refer to equation~\ref{eq:vel}).

The strategies are tested on both functions by performing 1,000 independent optimizations runs with random initial guesses distributed in the searchable parameter space bounded by [-100:100] for all dimensions. In the case of LM, the searchable parameter space is not bounded. This did not influence its performance as the OFs of both functions have generally increasing trends towards the boundaries (Figure~\ref{fig:tests_3D}). Optimization success is defined as identifying a solution with all parameters values within 0.1 of the global minimum parameter values ($\mathbf{x} = \mathbf{1}$ for the Rosenbrock function and $\mathbf{x} = \mathbf{0}$ for the Griewank function). The maximum number of function (model) evaluations ($E_{max}$) for the strategies is set to 20,000. However, in performed analyses, LM runs terminate at fewer function evaluations as the convergence criteria of LM are designed to terminate its run once it identifies a minimum in the response surface. The ability of LM to identify the global minimum depends on whether the minimum encountered by LM is local or global. 

\begin{figure}
	\begin{center}
		\includegraphics[width=10cm]{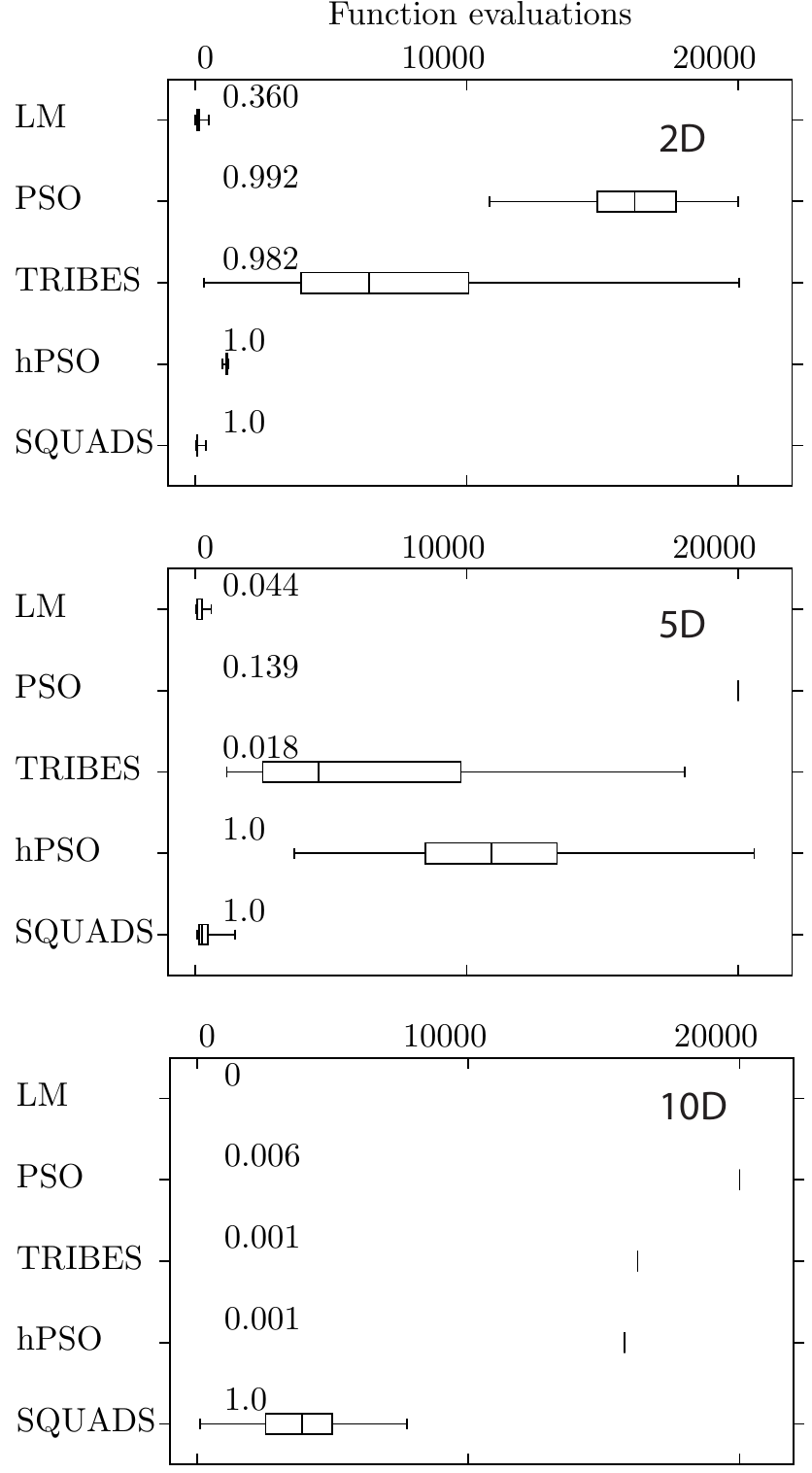}
	\caption{Boxplots of number of function evaluations to reach the global minimum for the 2D, 5D, and 10D Rosenbrock function. The boxes represent the 25$^{\mbox{th}}$ to 75$^{\mbox{th}}$ percentile ranges, the bars inside of the boxes represent the median values, and the whiskers represent the min and max values. Note that the statistical definitions are not accurate when the number of successful runs does not present a statistically significant sample. The fraction of successful runs out of 1000 for each strategy is stated above the boxes. The maximum allowable function evaluations for each run is 20,000.}
	\label{fig:box_r}
	\end{center}
\end{figure}

\begin{figure}
	\begin{center}
		\includegraphics[width=10cm]{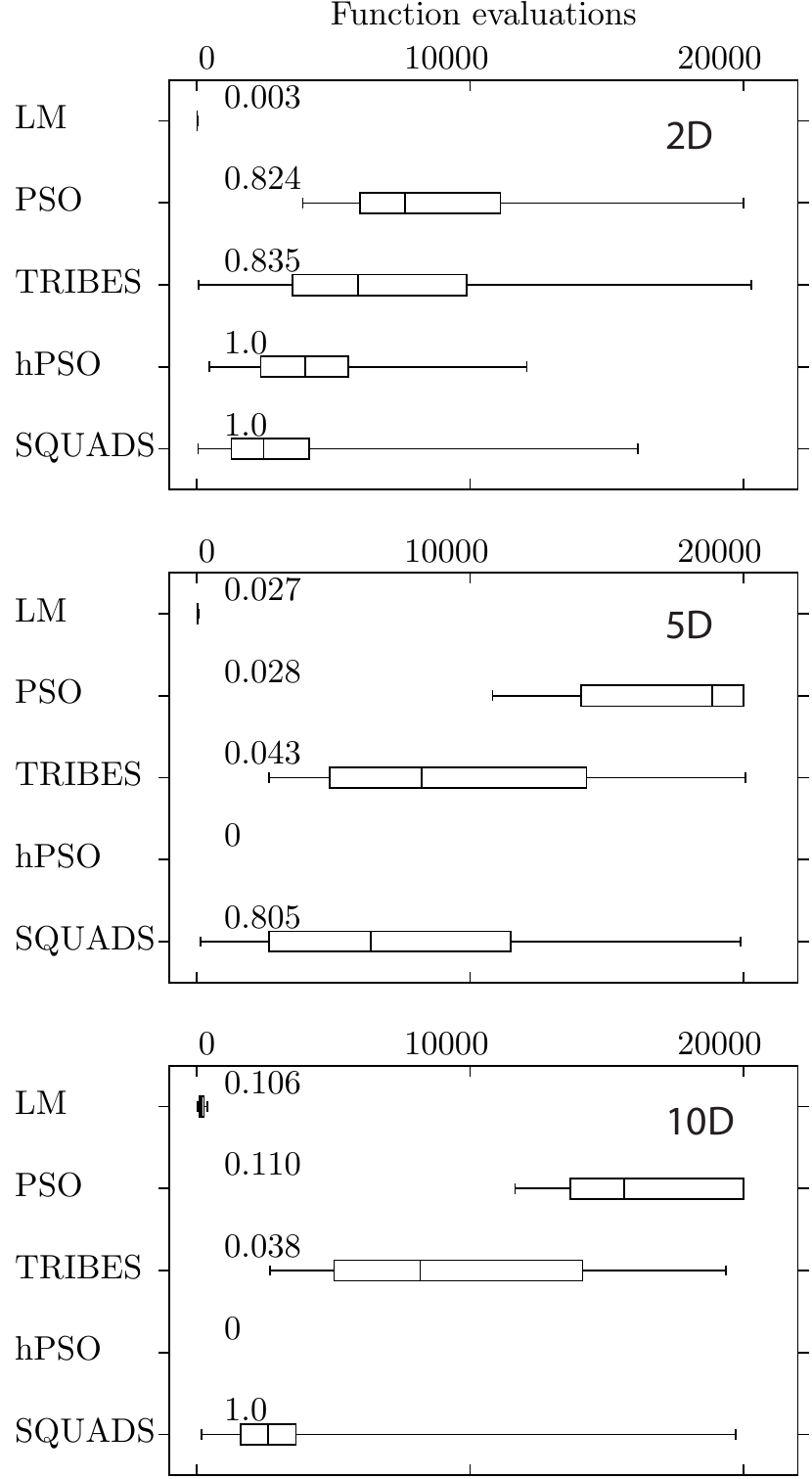}
	\caption{Boxplots of number of function evaluations to reach the global minimum for the 2D, 5D, 10D Griewank function. The boxes represent the 25$^{\mbox{th}}$ to 75$^{\mbox{th}}$ percentile ranges, the bars inside of the boxes represent the median values, and the whiskers represent the min and max values. Note that the statistical definitions are not accurate when the number of successful runs does not present a statistically significant sample. The fraction of successful runs out of 1000 for each strategy is stated above the boxes. The maximum allowable function evaluations for each run is 20,000.}
		\label{fig:box_g}
	\end{center}
\end{figure}

Figures~\ref{fig:box_r} and \ref{fig:box_g} present boxplots for the number of function evaluations for successful runs for 2D, 5D, and 10D Rosenbrock and Griewank functions, respectively. In the figures, the boxes represent the 25$^{\mbox{th}}$ to 75$^{\mbox{th}}$ percentile ranges, the bars inside of the boxes represent the median values, and the whiskers represent the minimum and maximum values. The fraction of successful runs out of the attempted runs are presented above the boxes. Note that the statistical definitions of the boxplots are not accurate for the cases where the number of successful runs does not present a statistically significant sample. The robustness of the strategies is defined as the percentage of successful runs (i.e.\ fraction of successful runs * 100). The efficiency of the strategies is summarized by the statistics presented in the boxplots.

For the Rosenbrock function (Figure~\ref{fig:box_r}), the robustness of LM decreases from the 2D case to the 10D case from 36\% to 0\%. The robustness of PSO and TRIBES is comparable in the 2D case, albeit with TRIBES exhibiting higher efficiency in general. In the 5D case, PSO has a higher robustness than TRIBES, however, at lower efficiency. hPSO achieves 100\% robustness in the 2D and 5D cases, with a significant decrease in efficiency from the 2D to 5D case. The robustness of hPSO decreases significantly in the 10D case with only a single success out of 1000 (0.1\%). In the 10D case, LM, PSO, TRIBES, and hPSO exhibit low robustness. \emph{Squads} is 100\% robust in all cases. The efficiency is observed to decrease from the 2D case to the 10D case for \emph{squads}; however, the efficiency of \emph{squads} is greater than PSO, TRIBES, and hPSO in all cases. The efficiency of \emph{squads} and LM are similar for the 2D and 5D cases (Figure~\ref{fig:box_r}). However, in these two cases, the robustness of \emph{squads} is 100\% which is considerably better than the robustness of LM (36\% for 2D and 4\% for 5D). In the 10D case, LM did not produce a single successful run while \emph{squads} is still 100\% robust.

For the Griewank function (Figure~\ref{fig:box_g}), as expected (see \cite{Locatelli03}), the robustness of LM increases as the dimensionality of the problem increases. In the 2D case, which is the most difficult for a local gradient-based strategy \citep{Locatelli03}, the robustness is only 3\%. Since LM is a local strategy, it is not surprising that LM frequently converges at non-optimal minima. As expected for the 2D case, the global strategies (PSO, TRIBES, hPSO and \emph{squads}) are substantially more robust than LM. The robustness of PSO and TRIBES (both purely global strategies) decrease significantly from the 2D to the 5D case, while decreasing only slightly from the 5D to the 10D case (the efficiency of PSO decreases also). hPSO is 100\% robust for the 2D case; however, is unable to locate the global minimum in the 5D and 10D cases. \emph{Squads} is 100\% robust in the 2D and 10D cases and 80\% robust in the 5D case.

As already discussed, the multidimensional Griewank function is important for testing of hybrid strategies such as \emph{squads}. For different parameter-space dimensionality, the performance of \emph{squads} is influenced by the ability of the adaptive rules in the optimization algorithm to balance between the local (LM) and global (APSO) strategies. With the increase of dimensionality, the local gradient-based (LM) strategy becomes more robust, while the global (APSO) strategy becomes less robust. At $D=2$, \emph{squads} is both more robust and efficient than the other global methods (Figure~\ref{fig:box_g}). At $D=10$, \emph{squads} benefits from the local gradient-based search strategy which performs better at higher dimensions. The 5D Griewank function is observed to be the most difficult test problem for \emph{squads} as both the local and global strategies struggle in this dimensionality of the Griewank function. Nevertheless, for the 5D case, \emph{squads} produces the highest robustness (80\%) and efficiency (excluding LM) of all the tested strategies; \emph{squads} is 100\% robust if the maximum number of function evaluations is increased to 70,000 (results are not shown here). In summary for the Griewank cases, \emph{squads} is observed to have the best performance when both robustness and efficiency are taken into consideration than the other strategies (Figure~\ref{fig:box_g}).

\begin{figure}
	\begin{center}
		\includegraphics[width=8cm]{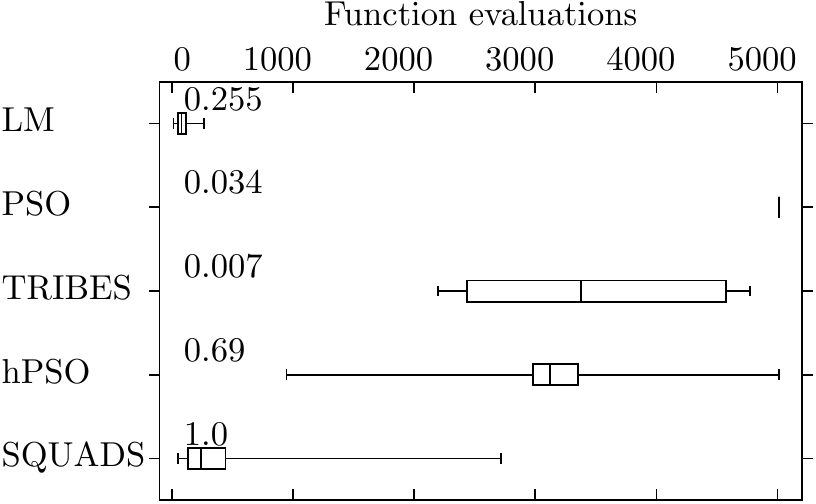}
		\caption{Boxplots of number of function evaluations to reach the global minimum for the hydrogeologic application presented in Section~\protect\ref{sect:cont}. The boxes represent the 25$^{\mbox{th}}$ to 75$^{\mbox{th}}$ percentile ranges, the bars inside of the boxes represent the median values, and the whiskers represent the min and max values. Note that the statistical definitions are not accurate when the number of successful runs does not present a statistically significant sample. The fraction of successful runs (out of 1000 for LM, PSO, TRIBES, and \emph{squads}; out of 100 for hPSO) is stated above the box. The maximum allowable function evaluations for each run is 5000.}
		\label{fig:cont_box}
	\end{center}
\end{figure}

The performance of the strategies is demonstrated on the hydrogeologic application in contaminant source identification presented in Section~\ref{sect:cont}. A boxplot of the necessary function evaluations for successful runs is presented in Figure~\ref{fig:cont_box}. As with the test functions, 1000 runs are performed for each strategy, except for hPSO, where only 100 runs are performed due to the computational expense of evaluating the contaminant transport model from hPSO (\~2.2 function evaluations per second). The maximum number of function evaluations allowed for each optimization run is limited to 5,000. An optimization run is considered successful if the objective function is reduced below a value of 1. This ensures that the solution has reached the area of attraction around the global minimum as the suboptimal minima of the response surface are all greater than 1. As with the test functions, it is observed that LM is efficient on the hydrogeologic application, but only approximately 26\% robust. PSO and TRIBES are observed to be inefficient and not robust in this case requiring high numbers of function evaluations with approximately 3\% and 0.7\% robustness,respectively. hPSO demonstrates some robustness at 69\%, but with low efficiency in general with a large variability in the necessary number of function evaluations. \emph{squads} demonstrates high robustness at 100\% with higher efficiency than PSO, TRIBES, and hPSO. While the efficiency of LM is better than \emph{squads} in this case, this is with a significantly lower robustness.

It is important to emphasize in all test cases, \emph{squads} can converge at a relatively low number of model evaluations when compared to PSO, TRIBES and hPSO. This is manifested by the minimum values of the boxplots in Figures~\ref{fig:box_r}, \ref{fig:box_g}, and \ref{fig:cont_box}. Furthermore, the statistical distributions of the number of model evaluations required to achieve the global minimum for \emph{squads} are skewed to the left in all cases (Figures~\ref{fig:box_r}, \ref{fig:box_g}, and \ref{fig:cont_box}). This demonstrates that more frequently \emph{squads} may converge with lower number of functional evaluations.

\section{Conclusions}

A new adaptive global hybrid optimization strategy called \emph{squads} is developed for solving computationally intensive inverse problems involving models representing the behavior of complex systems. \emph{Squads} utilizes a (1) global strategy for robust exploration of the parameter space to identify multiple areas of attraction and (2) a local gradient-based search strategy to efficiently locate the optimum  of areas of attraction. In essence, \emph{squads} is a global strategy that implements a local gradient-based optimization speedup. \emph{Squads} is sufficiently robust in avoiding becoming stuck in local minima during the optimization as typically observed in the case of local gradient-based strategies. The new strategy reduces the number of model runs typically required of other frequently used global strategies such as Particle Swarm Optimization (PSO) by efficiently exploring local areas of attraction. 

The strategy is tested on 2D, 5D, and 10D variations of two commonly used polynomial test functions: the Rosenbrock and Griewank functions. The strategy is also demonstrated on a synthetic hydrogeologic application to identify the source center and source dimensions of a contaminant plume in an aquifer based on observed contaminant concentrations at monitoring wells. The robustness of a strategy is defined as the percentage of runs that identify the global minimum in each test case. The efficiency of a strategy is evaluated through a statistical representation of the number of function or model evaluations necessary to identify the global optimum. In all cases, \emph{squads} is as robust or more robust than the other tested strategies: LM, PSO, TRIBES, and hPSO. \emph{Squads} is more efficient than PSO, TRIBES, and hPSO in all cases. For the Rosenbrock function, \emph{squads} has comparable efficiency to LM, however, in these cases, the robustness of \emph{squads} (100\%) is considerably better than the robustness of LM (less than 36\%). For other optimization problems, \emph{squads} may converge for the same number of functional evaluations as LM (Figure~\ref{fig:box_g}). For the Griewank function and hydrogeologic application, LM successfully converges to local areas of attraction; however, these were not always the global minimum. For the 2D Rosenbrock function, LM converges within an area of attraction shaped as a narrow curved valley where the gradient is so small that it appears to LM that it has identified a minima (the valley contains the global minimum).

The application of the \emph{squads} strategy is performed using the code MADS \citep{MADS10}. MADS and other files needed to execute the synthetic problems presented in this paper are available at http://www.ees.lanl.gov/staff/monty/codes/mads.html

\bibliographystyle{authordate1} 
\bibliography{squads} 

\end{document}